\documentstyle[twocolumn,pre,aps]{revtex}
\begin{document}
\draft
\title{Extended series expansions
for random sequential adsorption}
\author{Chee Kwan Gan$\dag\ddag$ and Jian-Sheng Wang$\dag$}

\address{
$\dag$Department of Computational Science, \\
National University of Singapore, Singapore 119260,\\
Republic of Singapore. \\
$\ddag$Institute of Materials Research and Engineering,\\
10 Kent Ridge Crescent, Singapore 119260, \\
Republic of Singapore.
}

\date{23 September 1997, Revised 29 October 1997}

\maketitle

\begin{abstract}

We express the coverage (occupation fraction) $\theta$, in powers of
time $t$ for four models of two-dimensional lattice random sequential
adsorption (RSA) to very high orders by improving an algorithm
developed by the present authors [J. Phys. A {\bf 29,} L177 (1996)].
Each of these series is, to the best of our knowledge, the longest at
the present.  We analyze the series and deduce accurate estimates for
the jamming coverage of the models.

\end{abstract}

\pacs{PACS number(s): 05.50.+q, 05.20.-y, 05.70.Ln}

\section{Introduction}

A large variety of physical, chemical, biological, and ecological
processes can be modeled by random sequential adsorption (RSA)
\cite{Evans-93} on a lattice. In this irreversible process,
particles are deposited randomly on a surface one after another.  In
the simplest form, the depositing particles are represented by
non-overlapping hard core extended objects.  These particles, once
adsorbed on a surface, are assumed to be fixed in their spatial
positions and exclude certain regions from further occupation.  It is
therefore interesting to study the coverage (fraction of occupied
sites) of the system as time evolves, in particular the jamming
coverage, a standstill beyond which no deposition is possible.

Exact solutions for RSA models are usually restricted to special cases
only \cite{exact-39,Fan-Percus-91}.  For those models which resist an
exact theoretical treatment, different approximate methods have been
developed \cite{Evans-93}. Series expansion is one of the powerful
analytical methods in RSA studies
\cite{Baram-Kutasov-89,Dickman-etal-91,%
Oliveira-etal-92,Bonnier-etal-93,Baram-Fixman-95,%
Gan-Wang-96,Gan-Wang-97}.  In this work we present an improved
algorithm for series expansions based on hierarchical rate equations
\cite{Gan-Wang-96,Gan-Wang-97}.  It is then applied to four lattice
RSA models where we are able to obtain, to the best of our knowledge,
the longest series known for each of model that we have studied.

\section{Models}
\label{sec:model}
Four models are studied in this work. For each of these models, we
start with an initially empty, infinite lattice. Particles are dropped
randomly and sequentially at a rate of $k$ per lattice site per unit
time, onto the lattice. Hereafter we set $k$ equal to unity without
loss of generality.  In the first model, monomers are deposited on a
square lattice with nearest-neighbor (NN) exclusion, where an adsorbed
monomer excludes its occupied sites and its four nearest-neighbors
from further occupation
\cite{Dickman-etal-91,Baram-Fixman-95}. In 
the second model, dimers of random orientations are deposited on a
square lattice.  In this model, a dimer can occupy a pair of
nearest-neighbor sites only if both of the chosen sites are vacant
\cite{Gan-Wang-96,Nord-Evans-85}.
The third and the fourth models are defined similarly as the
first and the second models, respectively, except that depositions
are performed on a hexagonal lattice
\cite{Nord-Evans-85,Caser-Hilhorst-94}.

\section{Modified algorithm}
\label{sec:algo}
\unitlength=6pt
\def\ck{\circle{0.6}}
\def\oA{\begin{picture}(1,1)(-0.5,-0.5)  
\put(0,0){\ck}
\end{picture}}
We shall first summarize the original algorithm presented in Ref
\cite{Gan-Wang-96}. We are interested in expressing the coverage
$\theta$ as a function of time $t$, which is related to the
probability of finding a single vacant site (denoted by `$\oA$'). More
precisely, we write $\theta(t) = 1 - P(\oA, t)$.  In general we are
interested in expressing the probability of finding any particular
configuration of interest, $C_0$, which we write it as $P_{C_0} \equiv
P(C_0)$, as a power series expansion of time $t$, with the expansion
point $t=0$. We write $P_{C_0} (t) = \sum_{n=0}^\infty P_{C_0}^{(n)}
t^n/n!$, with $n$th derivative of $P_{C_0}$ given by $P_{C_0}^{(n)} =
\left. \frac{d^n P_{C_0} (t) }{dt^n} \right|_{t=0}$.

Let $G_i$ denote the set of new configurations generated in the
calculation of the $i$th derivative of $P_{C_0}$, and $G_i^j$ the
corresponding $j$th derivatives of the set of configurations.  We
observe that $G_0^{n-1}$, $G_1^{n-2}$, $\ldots$, $G_{n-1}^0$
[determined at the $(n-1)$th derivative], $G_0^{n-2}$, $G_1^{n-3}$,
$\ldots$, $G_{n-2}^0$ [determined at the $(n-2)$th derivative],
$\ldots$, $G_0^0$ are predetermined before calculating the $n$th
derivative of $P_{C_o}$. In the calculation of $n$th derivative of
$P_{C_o}$, we determine systematically $G_0^n$, $G_1^{n-1}$, $\ldots$,
$G_{n-1}^1$, $G_{n}^0$, by recursive use of rate equations. This
algorithm is efficient since each value in $G_i^{n-i}$, $ 0 \le i \le
n$ and the rate equation for each configuration $C$ are generated once
only.  However, this algorithm consumes the memory quickly due to
storage of intermediate results. In this algorithm, each configuration
is transformed into a unique canonical representation and stored as
such in memory. If a configuration is already generated before, only a
pointer reference to the previous configuration is made. The checking
of the existence of a particular configuration is done efficiently
through the hashing technique \cite{Cormen-etal-89}.

An instructive way of looking at this algorithm is through the concept
of tree traversal \cite{Cormen-etal-89}.  We make the first
configuration $C_0$, of which its highest order of derivative to be
obtained is $h$, as the root node of the tree. The depth at which the
root node resides is set arbitrarily to $0$, this corresponds to the
index $i = 0$ in the symbol $G_i^j$ introduced above.  We see that a
complete breadth-first tree traversal will introduce at new set of
leaves nodes of depth which is one more than the depth of leaves nodes
in the previous breadth-first traversed tree.  Therefore at the end of
$D$th times of tree traversal we will introduce a set of new leaves
nodes of depth $D$.  Since we have to store all new configurations or
nodes that appear in the process, the breath-first tree traversal is
thus a very memory demanding algorithm.  To curb the exponential
memory growth, we stop the breadth-first traversal of tree when the
leaves nodes have a depth $ D'$ where a new and different strategy is
followed.  Notice now that the problem of finding the $h$th order
derivative of $C_0$ amounts to predetermining the first, second,
$\ldots$, through $(h-D')$ order derivatives for all the leaves nodes
of depth $D'$. If we can do that, then by another $(h-D')$ times of
breadth-first traversal of the tree from the root node $C_0$ again, we
can calculate all derivatives of $C_0$ up to order $h$. Now the
problem is to calculate the first through $(h-D')$th order derivatives
for each of the leaves nodes of depth $D'$. This can be achieved by
treating each of the leaves nodes of depth $D'$ as a {\it new} root
node in another breath-first tree traversal subproblem.  Each of these
subproblems can be dealt with one at a time, which means that we can
use the memory allocated for a subproblem for the next subproblem. It
is through this way that the use of memory is kept at roughly a fixed
amount. A drawback of this approach is that the derivatives for some
configurations have to be calculated more than once, however there is
an important observation that the calculations of all these
subproblems can be parallelized.

As another effort to cut the growth of the memory requirement, we
observe that the storing of the nodes is not really necessary for
nodes which have an absolute depth very close to $h$. This can be
achieved by using a depth-first tree traversal. When we are
considering the rate equation of a particular configuration $C$, we
recursively traverse down the tree whenever a configuration belongs to
the right hand side of the rate equation for node $C$, which we shall
call it a child node of $C$, appears. The merit of this depth-first
approach is that we do not need to consider all symmetry operations
that have to be applied to a configuration. These operations are
required before any configuration can be kept in the memory.  Although
the depth-first strategy effectively saves the troubles of finding a
canonical representation, the time to traverse a depth-first tree can
be quite long if the height of the tree is large.

A detailed description of the implementation of the original algorithm
can be found in Ref \cite{Gan-Wang-97}. It can be modified readily to
suit this new improved algorithm. Here we should mention the details
of the task of performing the rate equation expansion.  When the
canonical representation for a new configuration is required, we make
a list of 2-column vectors of entries which are the $x$ and $y$
coordinates of vacant sites specifying the configurations stored in a
``working matrix''.  Symmetry operations are applied to the new
configuration through pre-multiplying the vectors in the list with
appropriate transformation matrices.  Criteria are set so that we can
determine a canonical representation out of the list of new vectors.

Series expansions are parallelized using PVM version 3.3.11 \cite{PVM}
to utilize the hardware resources to the fullest. Calculations were
performed on a 16-node cluster of Pentium Pro 200 with a bonded dual
channel 100 BaseT Ethernet connection. All runs that we have done use
less than 60 Megabytes of main memory, with the longest run took an
elapsed time of 120 hours per machine in a dedicated computing
environment.  To compare the performance of this new algorithm and the
original algorithm \cite{Gan-Wang-96}, we notice that we are able to
obtain 4 more coefficients for the dimer RSA model on a square lattice
using the former algorithm.  The later algorithm requires 706
Megabytes of memory even at the 14th order derivative calculation.  The
results for all the models are displayed in Table
\ref{tab:sqr_ser_exp} and 
Table \ref{tab:hex_ser_exp} for square lattice RSA and hexagonal
lattice RSA, respectively. We note that our 18th order coefficient in
the expansion of powers of $u = e^{-t} - 1$, is $3560240252651011168$
for monomer RSA with NN exclusion on a square lattice, which is
slightly different from the value reported in Ref
\cite{Baram-Fixman-95}.

\section{Analysis of series}
\label{sec:ana_ser}

The approach to the jamming state for lattice RSA is often
exponential. This behavior enables us to use transformations of
variables which reflect the actual approach. The transformations that
we have employed are similar to that used in Refs
\cite{Dickman-etal-91,Gan-Wang-96}. First we transform the
coverage $\theta(t)$ in term of $y = 1-e^{-t}$. Another transformation
$z = (1-e^{-by})/b$ is performed. This transformation is suggested by
the fact that the exact jamming approach of dimer RSA on a linear
lattice corresponds to $\theta(t) = z$ with $b = 2$
\cite{Dickman-etal-91}. For other RSA
models, we set $b$ to be a free variable so that when different orders
of Pad\'e approximants are applied to $\theta(z)$, a crossing region
between different orders of Pad\'e approximants is to be located so as
to give a good estimate of jamming coverage $\theta_\infty$.  As an
example, Fig. \ref{fig:sqrnn_21.eps} shows the crossing region of
Pad\'e approximants of orders [11,10], [10, 11], [12, 9], [9, 12],
[13, 8], [8, 13], [14, 7], [7, 14], [10, 10], [11, 9], [9, 11], [12,
8], [8, 12], [10, 9], [9, 10], [11, 8], and [8, 11] for the series of
monomer RSA with NN exclusion on a square lattice, giving an estimate
of $\theta_\infty = 0.3641323(1)$, where the last digit denotes the
uncertainty.  This estimate is in good agreement with the estimate of
$\theta_\infty = 0.3641330(5)$ by Baram and Fixman
\cite{Baram-Fixman-95}. Analyzing the series 
using the square cactus as a reference model
\cite{Fan-Percus-91},
we obtain $ \theta_\infty = 0.364132(1) $. All these estimates from
the series analysis agree well with the simulation result of
$\theta_\infty = 0.36413(1)$ \cite{Meakin-etal-87}.

For the dimer RSA on a square lattice, we obtain $\theta_\infty =
0.906823(2) $. This is to be compared with the simulation results of
Oliveira {\it et al.}
\cite{Oliveira-etal-92} of
$\theta_\infty = 0.90677(6) $ and of Wang and Pandey
\cite{Wang-Pandey-96}
of $\theta_\infty = 0.906820(2) $. A somewhat biased estimate of
$\theta_\infty = 0.9068088(4) $ was obtained in Ref \cite{Gan-Wang-96}
as too small a range of $b$ was used.

For monomer RSA on a hexagonal lattice, $\theta_\infty =
0.37913944(1)$.  This result is much more accurate than the simulation
results of $\theta_\infty = 0.38(1)$ and $\theta_\infty = 0.379$
reported in Refs
\cite{Widom-66} and \cite{Evans-89}, respectively.
For the dimer RSA on hexagonal lattice, we obtain $\theta_\infty =
0.8789329(1)$. This is in good agreement with the estimate
$\theta_\infty = 0.87889$ in Ref \cite{Nord-Evans-85}.

\section{Conclusions}
\label{sec:summary}
In this work we have substantially extended the series for $\theta(t)$
by at least 3 terms for RSA models defined on a square lattice. The
series for the hexagonal lattice RSA are also of very high orders. We
noted that with the improvement of the original algorithm, very long
series and accurate estimates for $\theta_\infty$ can be obtained. The
generality of this computational method allows us to handle a variety
of problems based on rate equations. We are currently applying this
algorithm to provide more insights into the dynamics of the
two-dimensional kinetic Ising model
\cite{Wang-Gan-97}.

\section*{Acknowledgment}
This work was supported in part by an Academic Research Grant
No.~RP950601.

\bibliographystyle{plain}

\begin{thebibliography}{1}
\bibitem{Evans-93}
J. W. Evans, {Rev. Mod. Phys.} {\bf 65}, 1281 (1993).

\bibitem{exact-39}
P. J. Flory, {J. Am. Chem. Soc.} {\bf 61}, 1518 (1939);
J. J. Gonzalez, P. C. Hemmer and J. S. H$\o$ye, {Chem. Phys.} {\bf
3}, 228 (1974);
J. W. Evans, {J. Math. Phys.} {\bf 25}, 2519, 2527 (1984);
A. Baram and D. Kutasov, {J. Phys. A} {\bf 27}, 3683 (1994).

\bibitem{Fan-Percus-91}
Y. Fan and J. K. Percus, {Phys. Rev. Lett.} {\bf 67}, 1677 (1991).

\bibitem{Baram-Kutasov-89}
A. Baram and D. Kutasov, {J. Phys. A} {\bf 22}, L251
(1989).

\bibitem{Dickman-etal-91}
R. Dickman, J. -S. Wang, and I. Jensen,
{J. Chem. Phys.} {\bf 94}, 8252 (1991).

\bibitem{Oliveira-etal-92}
M. J.~de Oliveira, T. Tom\'e and R. Dickman, {Phys. Rev. A} {\bf 46}, 6294
(1992).

\bibitem{Bonnier-etal-93}
B. Bonnier, M. Hontebeyrie and C. Meyers, {Physica A} {\bf 198},
1 (1993).

\bibitem{Baram-Fixman-95}
A. Baram and M. Fixman, {J. Chem. Phys.} {\bf 103}, 1929 (1995).

\bibitem{Gan-Wang-96}
C. K. Gan and J. -S. Wang, {J. Phys. A} {\bf 29}, L177
(1996).

\bibitem{Gan-Wang-97}
C. K. Gan and J. -S. Wang, {Phys. Rev. E} {\bf 55}, 107 (1997).

\bibitem{Nord-Evans-85}
R. S. Nord and J. W. Evans, {J. Chem. Phys.} {\bf 82}, 2795 (1985).

\bibitem{Caser-Hilhorst-94}
S. Caser and H. J. Hilhorst, {J. Phys. A} {\bf 27},
7969 (1994).

\bibitem{Cormen-etal-89}
T. H. Cormen, C. E. Leiserson and R. L. Rivest, {\it Introduction
to Algorithms}, The MIT Press, McGraw-Hill Book Company (1989).

\bibitem{PVM} PVM (Parallel Virtual Machine) is a software package that
permits a heterogeneous collection of Unix computers hooked together
by a network to be used as a single large parallel computer.  See A.
Geist, A. Beguelin, J. Dongarra, W. Jiang, R. Manchek, V. Sunderam,
{\sl PVM: Parallel Virtual Machine, A Users' Guide and Tutorial for
Networked Parallel Computing}, MIT Press (1994).


\bibitem{Meakin-etal-87}
P. Meakin, J. L. Cardy, E. Loh, Jr., and D. J. Scalapino, {J. Chem.
Phys.} {\bf 86}, 2380 (1987).

\bibitem{Wang-Pandey-96}
J. -S. Wang and R. B. Pandey, {Phys. Rev. Lett.} {\bf 77}, 1773 (1996).

\bibitem{Widom-66}
B. Widom, {J. Chem. Phys.} {\bf 44}, 3888 (1966).

\bibitem{Evans-89}
J. W. Evans, {Phys. Rev. Lett.} {\bf 62}, 2642 (1989).

\bibitem{Wang-Gan-97}
J. -S. Wang and C. K. Gan, in preparation (1997).


\end{thebibliography}

\newpage
\widetext

\begin{table}
\caption{Series-expansion coefficients for the probability of
finding a vacant site of RSA models on the square lattice, with
$n$ designating the order of derivative.}
\label{tab:sqr_ser_exp}
\begin{tabular}{rrrr}
$n$ & \quad monomer RSA with NN exclusion & dimer
RSA \\
\hline\hline
1  &  $-$1  & $-$4 \\  
2  &  5  & 28 \\  
3  &  $-$37  & $-$268 \\  
4  &  349  & 3212 \\  
5  &  $-$3925  & $-$45868 \\  
6  &  50845  & 756364 \\  
7  &  $-$742165  & $-$14094572 \\  
8  &  12017245  & 292140492 \\  
9  &  $-$213321717  & $-$6653993260 \\  
10  &  4113044061  & 164952149516 \\  
11  &  $-$85493084853  & $-$4416119044972 \\  
12  &  1903886785277  & 126863203272268 \\  
13  &  $-$45187885535477  & $-$3889473277203116 \\  
14  &  1137973688508989  & 126677386324657804 \\  
15  &  $-$30289520203949205  & $-$4365431744153008620 \\  
16  &  849248887429012733  & 158621692698054953164 \\  
17  &  $-$25007259870924817749  & $-$6058617368871081964076 \\  
18  &  771322713104711008093  & \quad\quad\quad 242593022530588935132428 \\  
19  &  $-$24860884250598911650645  & \\  
20  &  835568036857675195155997  & \\  
21  &  $-$29227671255970830546587445 & \\  
\end{tabular}
\end{table}

\begin{table}
\caption{Series-expansion coefficients for the probability of
finding a vacant site of RSA models on the hexagonal lattice, with $n$
designating the order of derivative.}
\label{tab:hex_ser_exp}
\begin{tabular}{rrrr}
$n$ & \quad monomer RSA with NN exclusion & dimer
RSA \\
\hline\hline
1  &  $-$1  & $-$3 \\  
2  &  4  & 15 \\  
3  &  $-$22  & $-$99 \\  
4  &  154  & 807 \\  
5  &  $-$1306  & $-$7803 \\  
6  &  12946  & 87039 \\  
7  &  $-$146026  & $-$1097139 \\  
8  &  1837666  & 15383607 \\  
9  &  $-$25429018  & $-$237018699 \\  
10  &  382667218  & 3973893519 \\  
11  &  $-$6208467946  & $-$71934898755 \\  
12  &  107847476914  & 1396766151303 \\  
13  &  $-$1994552336218  & $-$28932596959515 \\  
14  &  39089479606162  & 636355373676831 \\  
15  &  $-$808620801445066  & $-$14801609728262739 \\  
16  &  17597107973050354  & 362819105840203479 \\  
17  &  $-$401684435661166234  & $-$9343388660601611115 \\  
18  &  9593278364964882706  & 252096661713936722415 \\  
19  &  $-$239168310954693706954  & $-$7109230409530804525155 \\  
20  &  6211745800773276251122  & 209085655917886138504551 \\  
21  &  $-$167765099068145498846842  & $-$6400562190944268227677947 \\  
22  &  4703750163644209363257538  & \quad\quad 203577552403910264228987775 \\  
23  &  $-$136703709388456354551539146  & \\  
24  &  4112464853135639607791658706  & \\  
\end{tabular}
\end{table}

\clearpage
\widetext
\begin{figure}
\caption{Pad\'e approximant estimates for the jamming coverage
$\theta_\infty$ as a function of the transformation parameter $b$, for
the monomer RSA with NN exclusion on a square lattice.}
\label{fig:sqrnn_21.eps}
\end{figure}
\input epsf
\vfill
\epsfxsize\hsize\epsfbox{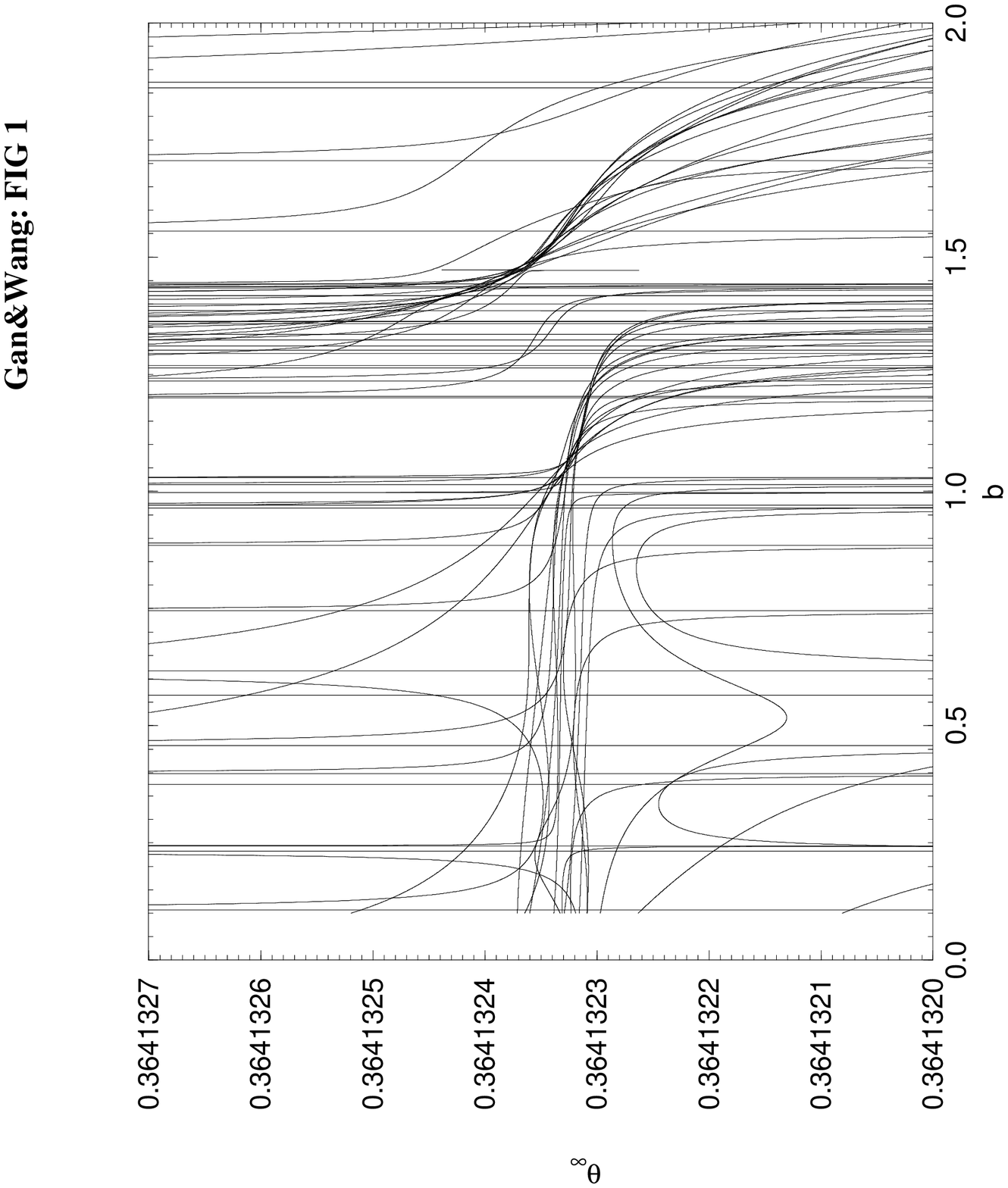}
\vfill
\end{document}